\documentclass{cimento}
\usepackage{graphicx,color,epsfig,epsf,rotating,caption}
\title{Theoretical issues on the top mass reconstruction\\ at hadron colliders}
\author{Gennaro Corcella\from{i}\from{j}\from{k}}

\instlist{ 
	\inst{i}  Museo Storico della Fisica, Centro Studi e Ricerche E.~Fermi,\\
             Piazza del Viminale 1, I-00184 Roma, Italy.   
\inst{j}Scuola Normale Superiore, Piazza dei Cavalieri 7, I-56123 Pisa, Italy.
\inst{k}INFN, Sezione di Pisa,  Largo Fibonacci 3,
I-56127, Pisa, Italy.}

\PACSes{
\PACSit{14.65.Ha }{Top quarks}\PACSit{12.38.Bx }{Perturbative calculations}}

\begin{document}
\maketitle
\begin{abstract}
I discuss a few selected topics related to the reconstruction of the 
mass of the top quark at hadron colliders. In particular, the
relation between the measured top mass and theoretical definitions,
such as the pole or $\overline{\rm{MS}}$ mass, is debated.
I will also summarize recent studies on the Monte Carlo uncertainty
due to the fragmentation of bottom quarks in top decays.
\end{abstract}

The top quark is an essential ingredient of the Standard Model of the
fundamental interactions and the determination of its properties provides
important tests of the strong and electroweak interactions.
In particular, the top quark mass ($m_t$) plays a crucial role in the
precision tests of the Standard Model, as the combination of $m_t$ with the
$W$-boson mass $m_W$ constrains the mass of the Standard Model Higgs boson
(see, e.g., the updated results from the LEP electroweak working 
group \cite{lep}).
Moreover, the top mass, along with $m_W$ and the mass of a possibly discovered
Higgs boson, is a relevant parameter to discriminate among 
New Physics scenarios, such as the Minimal Supersymmetric Standard Model
\cite{hey}. 
The latest top mass measurement at the Tevatron
accelerator is $m_t=(173.3\pm 1.1)$~GeV \cite{comb}.
However, it is often unclear how to relate
the measured $m_t$ to a consistent mass definition.

In fact, from the theoretical point of view, several 
mass definitions exist, according to how one subtracts the ultraviolet
divergences in the renormalized self-energy $\Sigma(p,m_t,\mu)$, where
$p$ is the momentum of the top quark, assumed to be off-shell,
and $\mu$ the renormalization scale.
%One thus constructs the renormalized propagator $S(p)$, whose inverse reads:
%\begin{equation}
%S^{-1}(p)=-i[\not p -m_t^0+\Sigma^R(p,m_t^0,\mu)].
%\end{equation}
The pole mass, often used in decay processes of on-shell
particles,
is thus defined by means of the conditions
$\Sigma(p)=0$ and $\partial\Sigma/\partial\hspace{-0.1cm}\not\hspace{-0.1cm} p=0$
%\begin{equation}
%\Sigma^R(p)=0 \ , \frac{\partial\Sigma^R}
%{\partial\hspace{-0.2cm}\not p}=0
%\end{equation}
for $\not \hspace{-0.1cm}p=m_t$.
The pole mass works well for leptons, such as electrons, whereas,
when dealing with heavy quarks,
it presents the so-called non-perturbative ambiguity, i.e. an
uncertainty $\Delta m\sim \Lambda_{\rm{QCD}}$.
In other words, after including higher-order corrections, the 
self-energy, expressed in terms of the pole mass $m_t$,
exhibits in the infrared regime a renormalon-like behaviour:
\begin{equation}
\Sigma(m_t)\sim m_t\ \sum_n\ \alpha_S^{n+1}\  
(2\beta_0)^n\ n!,
\end{equation}
where $\beta_0$ is the first coefficient of the QCD $\beta$-function.

Another definition of top mass which is often used is the 
$\overline{\rm{MS}}$ mass, namely $\bar m_t(\mu)$,
corresponding, e.g. in dimensional
regularization with $D=4-2\epsilon$ dimensions, to subtracting off the 
renormalized self-energy the
quantity $1/\epsilon+\gamma_E-\ln(4\pi)$, $\gamma_E$ being the
Euler constant.
The $\overline{\rm{MS}}$ mass is a suitable mass definition
in processes where the top quarks are off-shell;
in fact, by means of this definition, one is able to reabsorb in
$\bar m_t(\mu)$ contributions
$\sim\ln (\mu^2/m_t^2)$, which are large if $\mu$ is taken of the
order of the hard scale $Q$ and $Q\gg m_t$.
For $t\bar t$ production at threshold, nonetheless, the $\overline{\rm{MS}}$ mass 
is not an adequate mass scheme, since it exhibits contributions
$\sim (\alpha_S/v)^k$, $v$ being the top quark velocity, which are
enhanced for $v\to 0$. Of course, one can always 
relate pole and $\overline{\rm{MS}}$
masses by means of a relation like \cite{kuhn}:
\begin{equation}
m_t=\bar m_t(\mu)\left[1+\alpha_S(\mu)c_1+\alpha_S^2(\mu)c_2+\dots\right],
\end{equation}
where the coefficients $c_i$ depend on $\ln[\bar m_t^2(\mu)/\mu^2]$.

Besides pole and $\overline{\rm{MS}}$ schemes,
other mass definitions for heavy quarks have been proposed.
For example, in the potential-subtracted (PS) mass a counterterm
$\delta m_t$ is constructed in such a way to remove the renormalon ambiguity 
in the infrared regime at the factorization scale $\mu_F$.
It is given by \cite{beneke}:
\begin{equation}
m_{t,\mathrm{PS}}(\mu_F)=m_t-\delta m_t(\mu_F),
\end{equation}
where the subtracted term reads
\begin{equation}
\delta m(\mu_F)=\frac{1}{2}\int_{|q|<\mu_F}{\frac{d^3 q}{(2\pi)^3}}
\tilde V(q),
\end{equation} 
$\tilde V(q)$ being the Fourier transform of the $t\bar t$ Coulomb
potential.
More generally, one can define several  
short-distance masses in terms of a parameter,
usually called $R$, corresponding, e.g., to the factorization scale 
$\mu_F$ or the $\overline{\rm{MS}}$ mass, so that the pole mass $m_t$
can be expressed in terms of a $R$-dependent renormalized 
mass and a counterterm
\cite{hoang1}:  
\begin{equation}\label{mtr}
m_t=m_t(R,\mu)+\delta m_t(R,\mu).
\end{equation}
In Eq.~(\ref{mtr}) $\delta m_t(R,\mu)$ can be expanded as a series of the strong
coupling constant, whereas $m_t(R,\mu)$ satisfies a renormalization-group
equation: 
\begin{equation}
\delta m_t(R,\mu)=R\sum_{n=1}^\infty\sum_{k=0}^n
a_{nk}\alpha_S^n(\mu)\ \ ,\ \ 
\frac{dm(R,\mu)}{d\ln\mu}=-R\gamma[\alpha_S(\mu)],
\end{equation}
with 
$\gamma[\alpha_S(\mu)]$ playing the role of the anomalous dimension 
\cite{hoang1}.

In order to make a statement on
the top quark mass which is extracted from the data, in principle 
one should compare
the measurement of a quantity depending on the top mass 
with a calculation which uses a consistent definition, e.g.
the pole or the $\overline{\rm{MS}}$ mass.
The D0 Collaboration \cite{d0} compared the measured total
$t\bar t$ cross section with the
calculations \cite{cacciari} and \cite{moch}.
Ref.~\cite{cacciari} computes
the $t\bar t$ cross section at
next-to-leading order (NLO) and resums
soft/collinear contributions in the next-to-next-to-leading 
logarithmic (NLL) approximation. The calculation \cite{moch} is NLO,
resums the Sudakov logarithms of the top velocity
$\alpha_S^4\ln^kv$, with $k\leq 4$, and includes Coulomb contributions
$\sim 1/v$ and $\sim 1/v^2$ to next-to-next-to-leading
order (NNLO). 
Comparing the experimental cross section with the formulas
in Refs.~\cite{cacciari,moch}, one can extract the top mass
employed in such calculations. Using the pole mass,
which is a reasonable assumption since top quarks are slightly above
threshold at the Tevatron, one obtains the following results \cite{d0}:
$m_t=171.5^{+9.9}_{-8.8}$~GeV, according to \cite{cacciari},
$m_t=173.1^{+9.8}_{-8.6}$~GeV when using \cite{moch}.

Furthermore, in Ref.~\cite{moch} the computation was carried out 
in the $\overline{\rm{MS}}$ renormalization scheme, and  
$\bar m_t(\bar m_t)$ was extracted from the comparison with the measured
cross section. Fig.~1, taken from Refs.~\cite{d0,moch} 
shows the main results of such a comparison.
\begin{figure}
\begin{center}\label{moch1}
\centerline{\resizebox{0.33\textwidth}{!}{\includegraphics{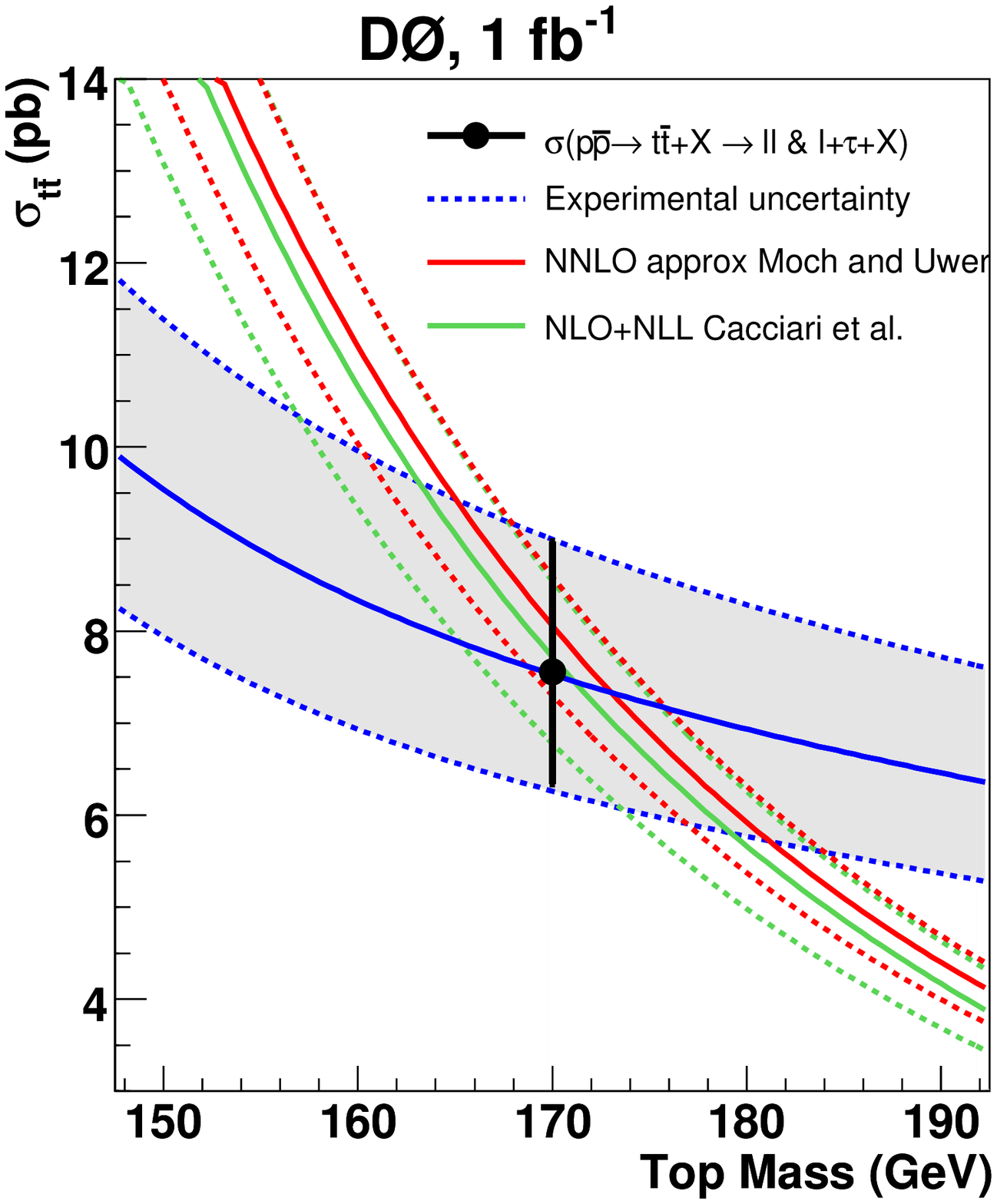}}%
\hfill%
\hspace{-1.cm}
\resizebox{0.42\textwidth}{!}{\includegraphics{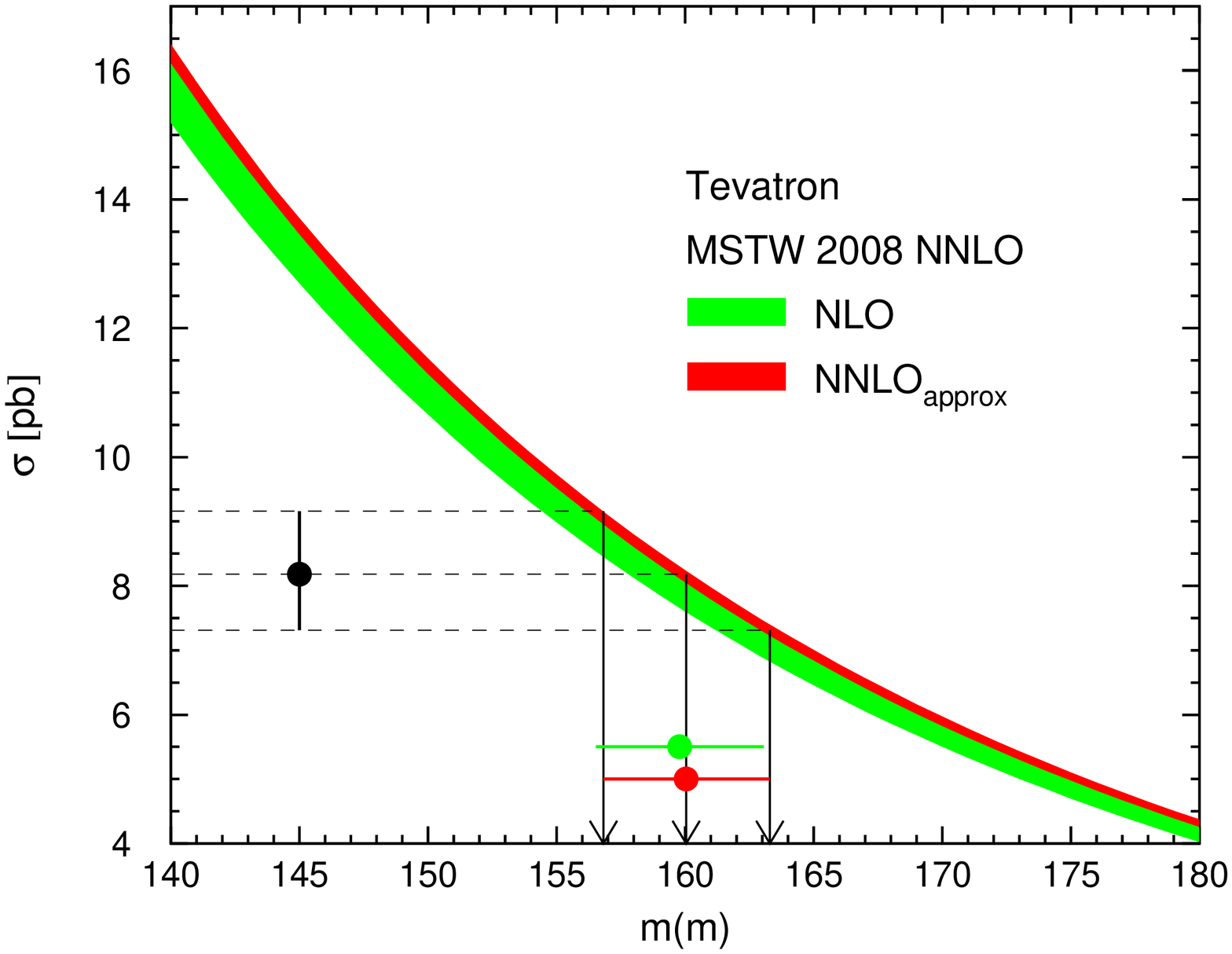}}}
\end{center}
\caption{Left: Extraction of the top pole mass by comparing the D0 cross section
with the computations \cite{cacciari,moch}. 
Right: Determination of the $\overline{\rm{MS}}$
mass following \cite{moch}.} 
\end{figure}
From $\bar m_t(\bar m_t)$ one can determine the pole mass
$m_t$: the central values are reported in
Table~\ref{tabmt}, whereas the errors are quoted in \cite{moch}. 
\begin{table}\small
\begin{tabular}
{|c|c|c|}\hline
  & $\bar m_t(\bar m_t)$ & $m_t$   \\
\hline
LO & 159.2 GeV &   159.2 GeV \\
 \hline
NLO & 159.8 GeV & 165.8 GeV \\
\hline
NLO+ NNLO approx. & 160.0 GeV & 168.2 GeV\\
\hline
\end{tabular}
\caption{Extracted values of the pole and $\overline{\rm{MS}}$ top mass
comparing the computation \cite{moch} with the D0 $t\bar t$ cross section.
See \cite{moch} for the errors on the quoted numbers.}
\label{tabmt}
\end{table}
Ref.~\cite{moch} also found out that using the $\overline{\rm{MS}}$ top mass 
leads to a milder dependence of the cross section on factorization and renormalization
scales, with respect to the pole mass. This result is indeed quite cumbersome,
since, in principle, at the Tevatron top quarks are almost at threshold; the 
possible full inclusion
of NNLO terms should probably shed light on the uncertainty 
when using pole or $\overline{\rm{MS}}$ masses.

Although the D0 determination of the top mass by using fixed-order and
possibly resummed calculations is surely very interesting, the standard 
Tevatron analyses \cite{comb}, such as the template or matrix-element methods, 
are driven by Monte Carlo parton shower 
generators, such as HERWIG \cite{herwig} and PYTHIA \cite{pythia}.
Such algorithms simulate multiple radiation in the soft/collinear approximation
and are possibly supplemented by matrix-element corrections to 
include hard and large-angle emissions.
Moreover, the yielded total cross
section is LO, whereas differential distributions are equivalent
to leading-logarithmic soft/collinear approximation, with the
inclusion of some NLLs \cite{cmw}.
The hadronization transition is finally implemented according to the cluster
model \cite{cluster} in HERWIG and the string model \cite{string} in
PYTHIA.

The template and matrix-element methods rely on the Monte Carlo
description of top decays. As discussed above, when looking at top decays
near threshold, the reconstructed top mass should be close to the pole mass and
the world average actually agrees, within the error ranges, with the
pole mass extracted from NLO cross section calculations.
However, as parton shower generators are not NLO computations, 
there are several uncertainties which affect the correspondence between
the mass which is implemented and any mass definition and renormalization
scheme. Monte Carlo algorithms 
neglect width and interference effects, but just factorize 
top production and decay; this is a reasonable
approximation as long as one sets cuts on the transverse energy of final-state
jets much higher than the top width $\Gamma_t$. Nevertheless, the quoted systematic
and statistical errors on the top mass \cite{comb} are 
competitive with the top width. Furthermore, top decays ($t\to bW$) in both
HERWIG and PYTHIA are matched 
to the exact tree-level $t\to bWg$ calculation \cite{corsey1,norb}, 
but the virtual corrections in
the top-quark self-energy, which one must calculate to consistently define
a renormalization scheme,
are included only in the soft/collinear approximation by means of the
Sudakov form factor. Also, other uncertainties are associated with the
flow of the colour of the top quarks and with the hadronization
corrections, since all measured observables are at hadron-level.

In principle, higher-order calculations are available even for top decays:
Refs.~\cite{mitov,mitov1} calculated the $b$-quark spectrum in top decay
at NLO and included NLL soft/collinear resummation in the
framework of perturbative fragmentation functions. Nevertheless,
such calculations are too inclusive to be directly used in the
Tevatron analyses; also, the results are expressed in terms 
of the $b$-quark ($B$-hadron) energy fraction in top rest frame,
which is a difficult observable to measure.
Ref.~\cite{melnikov} recently
computed several quantities relying on top decays at NLO using the top pole mass: 
a comparison of this calculation with Tevatron or LHC data, though not carried out
so far, 
should possibly provide another consistent determination of the pole mass.

An attempt to associate the Monte Carlo top mass with a formal mass definition
has been carried out in \cite{hoang1,hoang,scimemi,hoang2} in the framework of
Soft Collinear Effective Theories (SCET).
According to this approach, valid in the regime 
$Q\gg m_t\gg \Gamma_t\gg \Lambda_{\rm{QCD}}$, where $Q$ is the process hard scale, 
one can factorize the double-differential cross section in terms
of the top ($M_t$) and antitop ($M_{\bar t}$) invariant-mass squared as follows:
\begin{equation}\label{scet}
\frac{d\sigma}{dM_t^2dM_{\bar t}^2}=\sigma_0 H_Q
H_m
\int{d\ell^+d\ell^-}B_+\left(\ell^+,\Gamma_t\right)
B_-\left(\ell^-,\Gamma_t\right)
S(\ell^+,\ell^-).
\end{equation}
In Eq.~(\ref{scet}), $H_Q$ and $H_m$ are hard-scattering coefficient functions 
at scales $Q$ and $m_t$, $B_{\pm}$ are the so-called heavy-quark jet functions, 
describing the evolution of top quarks into jets, 
$S(\ell^+,\ell^-,\mu)$ is named soft function and is a non-perturbative fragmentation function,
depending on soft radiation and ruling the dijet and mass distributions.
When writing a factorization formula like in (\ref{scet}), large logarithms of the scale 
ratios, such as $\ln(Q/m_t)$ or $\ln(m_t/\Gamma_t)$, clearly arise: the jet function
has been resummed to NNLL in $e^+e^-$ annihilation, the soft one to NLL \cite{hoang}. 
\begin{figure}[b]
\centerline{\resizebox{0.4\textwidth}{!}{\includegraphics{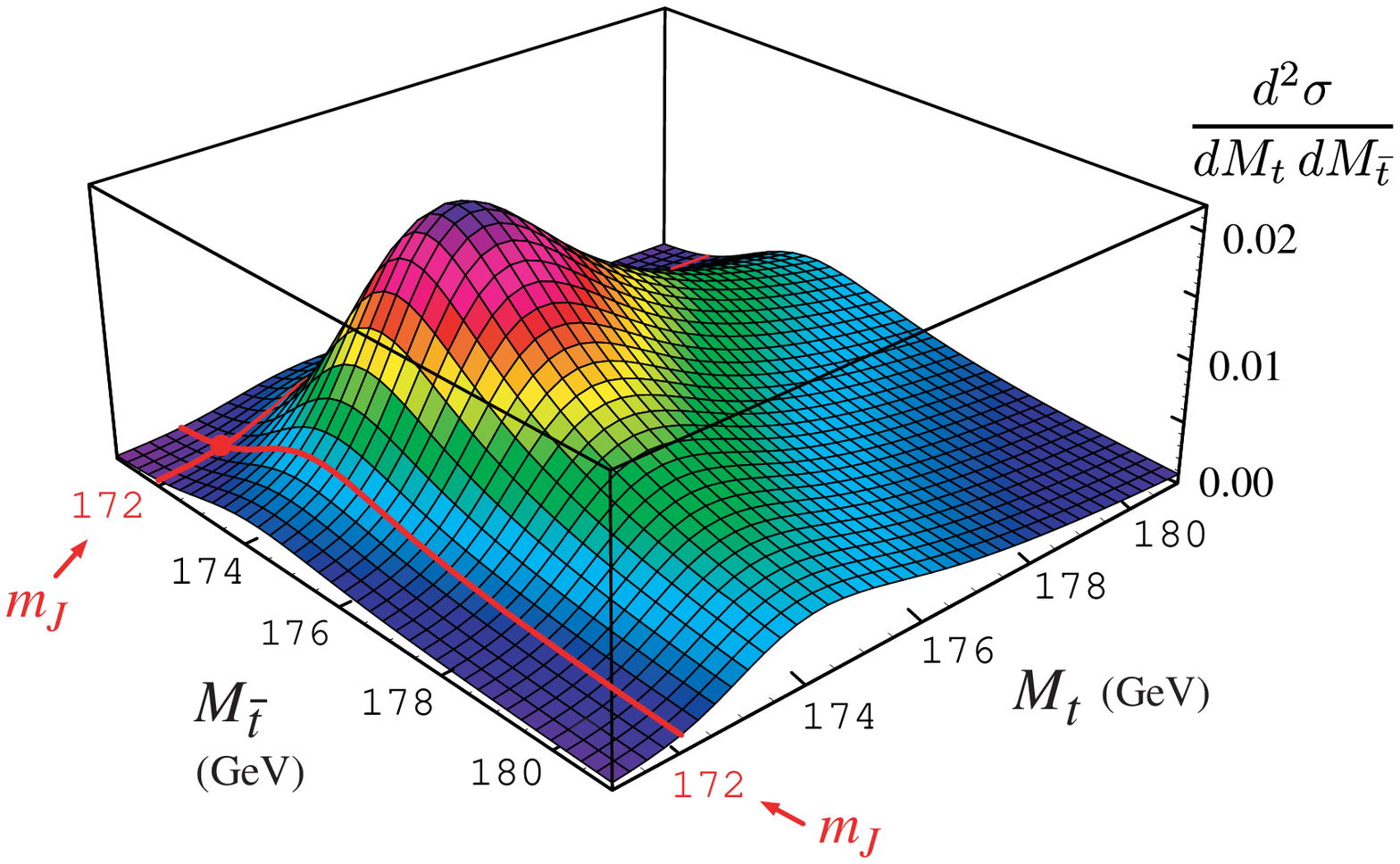}}%
\hfill%
\resizebox{0.4\textwidth}{!}{\includegraphics{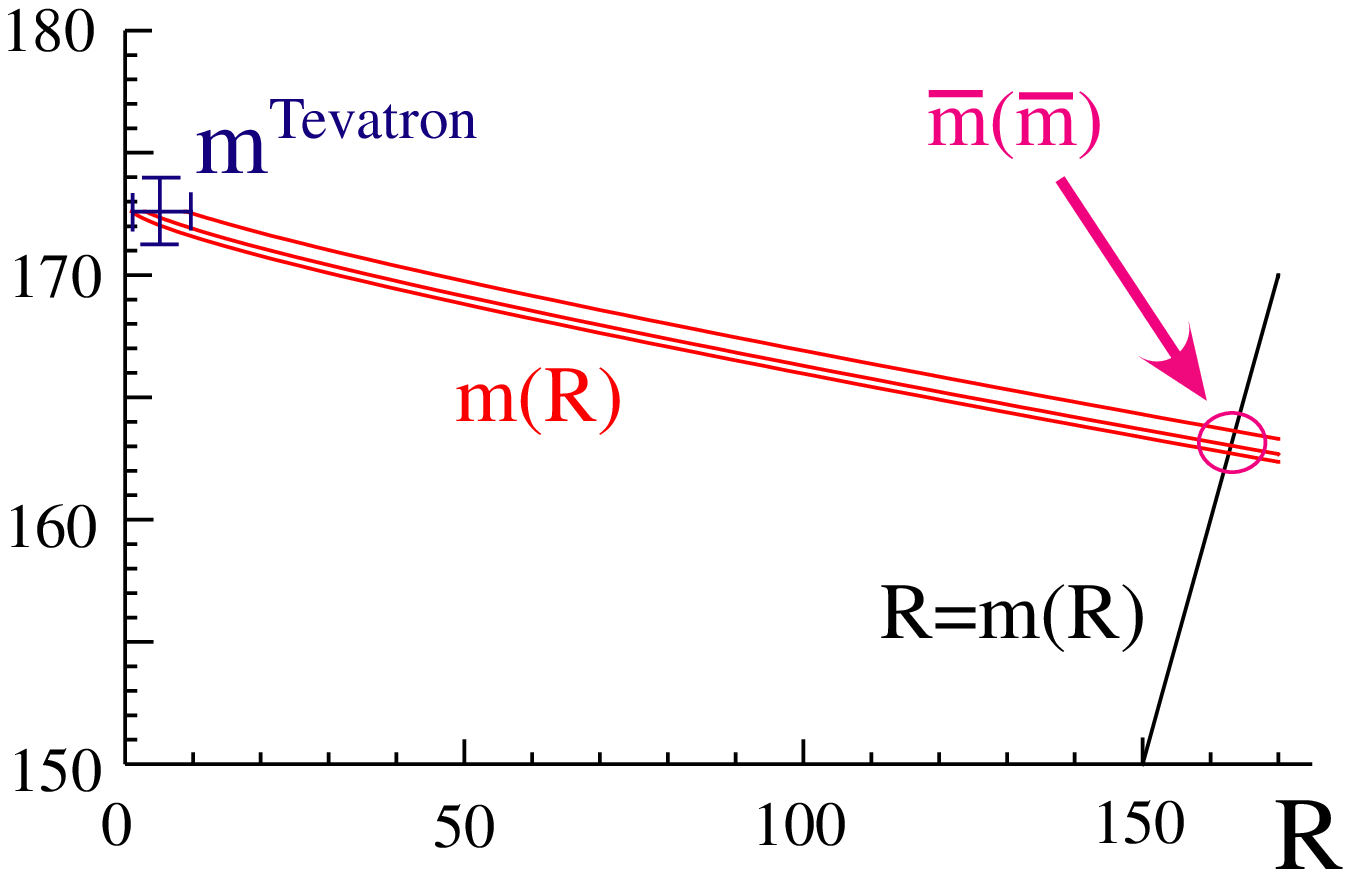}}}
\caption{Left: Double differential resummed cross section with respect to
the top and antitop invariant masses. Right: $R$-evolution from the
Tevatron top mass, corresponding to $R\simeq 
1$~GeV, to the $\overline{\rm{MS}}$ mass,
i.e. $R\simeq \bar m_t(\bar m_t)$.}
\label{figscet}\end{figure}
The peak value of the distribution $d^2\sigma/(dM_t^2dM_{\bar t}^2)$, 
displayed in Fig.~\ref{figscet} in the NLL approximation for 
$e^+e^-\to t\bar t$ processes, is independent of the mass scheme and 
can be expressed in terms of a short-distance mass $m$, $\Gamma_t$,
$Q$ and $\Lambda_{\rm{QCD}}$:
\begin{equation}
M_t^{\mathrm{peak}}=m+\Gamma_t(c_1\alpha_S+c_2\alpha_S^2+\dots)+
\frac{c_3Q\Lambda_{\rm{QCD}}}{m},
\end{equation}
where the term $\sim\Gamma_t$ depends on the jet function and the one 
$\sim\Lambda_{\rm{QCD}}$ on the soft function.
The jet mass is thus constructed 
as a short-distance mass, defined following Eq.~(\ref{mtr}),
with the parameter $R\sim \Gamma_t$ and the counterterm $\delta m_J$
depending on the jet function.
Expressing $M_t^{\mathrm{peak}}$ in terms of 
the pole ($m_t$) and jet ($m_{J}$) masses, one can finally relate $m_t$ and
$m_J(\mu)$ \cite{scimemi}:
\begin{equation}\label{jetmass}
m_t=m_{J}(\mu)+e^{\gamma_E}\Gamma_t
\frac{\alpha_S(\mu)C_F}{\pi}\left(\ln\frac{\mu}{\Gamma_t}
+\frac{1}{2}\right)
+{\cal O}(\alpha_S^2).
\end{equation}
One can note that a correction $\sim\Gamma_t$ arises in the difference
between the pole and jet masses: in fact, it was pointed out above that the 
top width was one of the ambiguities when associating the top mass reconstructed from
final-state decay observables, e.g. the jet mass, with the pole mass.
Ref.~\cite{hoang2} then assumed that the jet mass should correspond to the mass 
in the event generators and 
measured at the Tevatron with $\mu$ being of the order of the hadronization
scale (shower cutoff), i.e. $\mu\simeq Q_0\simeq 1$~GeV.
Using the $R$-evolution equations (\ref{mtr}), with $m_{J}(Q_0)\simeq$~172~GeV,
one can finally obtain the $\overline{\rm{MS}}$ mass 
$\bar m_t(\bar m_t)\simeq 163.0$~GeV \cite{hoang2}
(see Fig.~\ref{figscet}). 
Refs.~\cite{hoang1,scimemi,hoang2} seem therefore to indicate 
a possible strategy to relate the mass parameter in the Monte 
Carlo codes to pole and $\overline{\mathrm{MS}}$ masses.
However, as discussed also in \cite{hoang2}, the situation at hadron 
colliders is more complicated and,
before drawing a final conclusion, one would need to take into account 
effects, such as initial-state radiation, hadronization and underlying event,
which are included in Monte Carlo event generators, but not yet in the
factorization formula (\ref{scet}), which
has been worked out for $e^+e^-$ annihilation. 
Further studies aiming at extending the factorization (\ref{scet}) to
hadron colliders are certainly worth to be pursued.

In fact, the theoretical error on the top mass determination is a crucial
issue, even for the sake of the Standard Model precision tests.
Ref.~\cite{mescia} presents an analysis aimed at assessing the uncertainty due to
the treatment of bottom-quark fragmentation in top decays in HERWIG and PYTHIA.
For this scope, one should tune the cluster and string
models to the same data sets: Ref.~\cite{drol} found out that
the default parametrizations are unable to reproduce
LEP and SLD data on $B$-hadron production in $e^+e^-$ annihilation, but it
was necessary to fit the hadronization models to obtain an acceptable 
description of such data.
As in \cite{mescia}, an interesting observable
is the invariant mass
$m_{B\ell}$ in the dilepton channel, where $B$ is a $b$-flavoured
hadron in top decay and $\ell$ a lepton from $W$ decay $W\to\ell\nu$.
A reliable description of $m_{B\ell}$ is essential
in the study \cite{avto} (see also Chapter~8 in Ref.~\cite{cms1}), 
wherein the top mass is reconstructed by means
of the $m_{J/\psi\ell}$ spectrum, with the $J/\psi$ coming from 
$B\to J/\psi$ decays, and in the analysis 
\cite{lucio}, which investigates semileptonic decays
$B\to \mu X$ and
extracts the top mass from the $m_{\mu\ell}$ distribution.
Moreover, the $B$-lepton invariant mass was employed in \cite{cms} to 
gauge the impact of the implementation of
hard and large-angle radiation in the simulation of top decays; also, $m_{B\ell}$
is one of the observables calculated in the NLO approximation in Ref.~\cite{melnikov}.
\begin{figure}[t]
\centerline{\resizebox{0.4\textwidth}{!}{\includegraphics{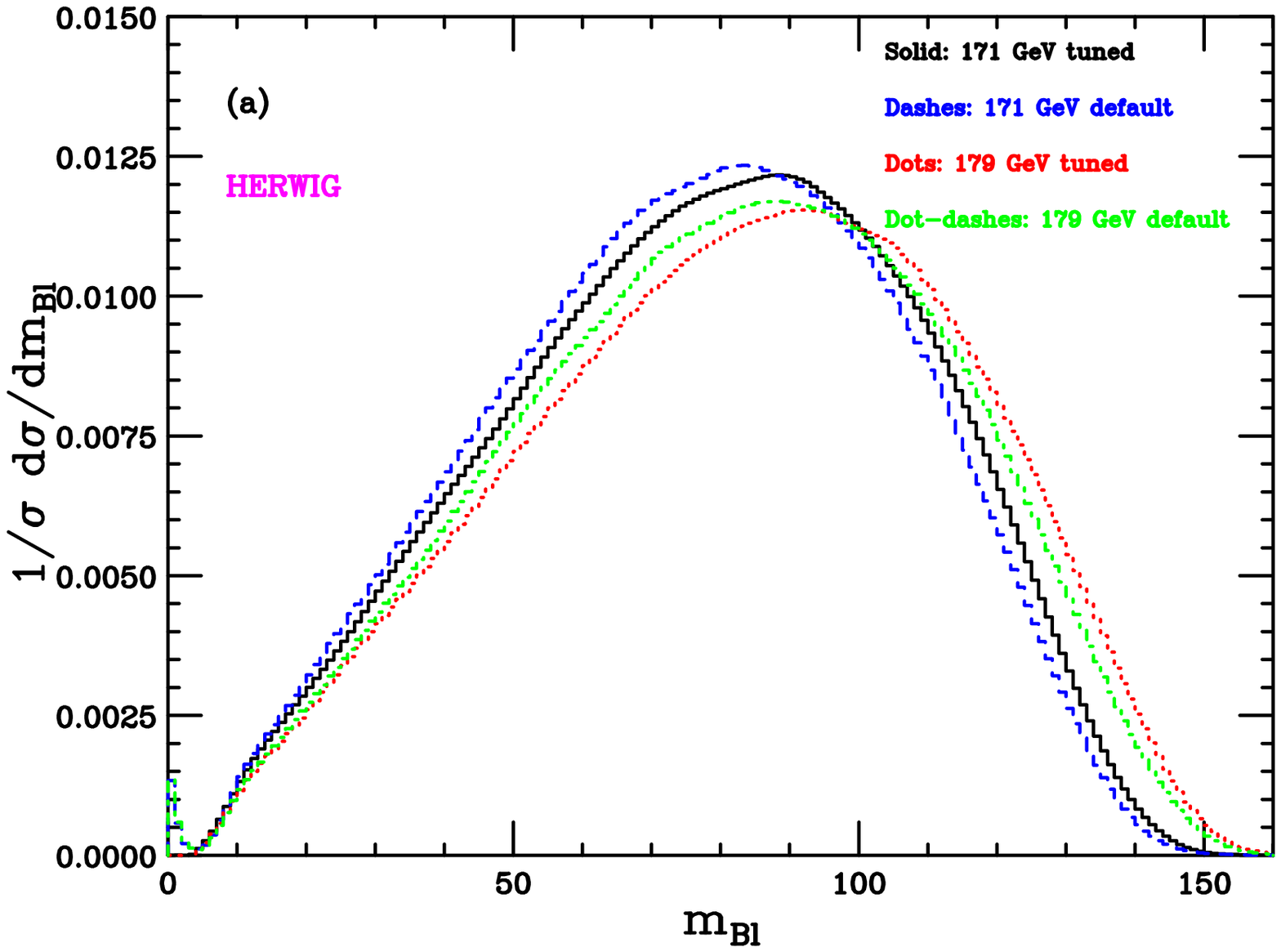}}%
\hfill%
\resizebox{0.4\textwidth}{!}{\includegraphics{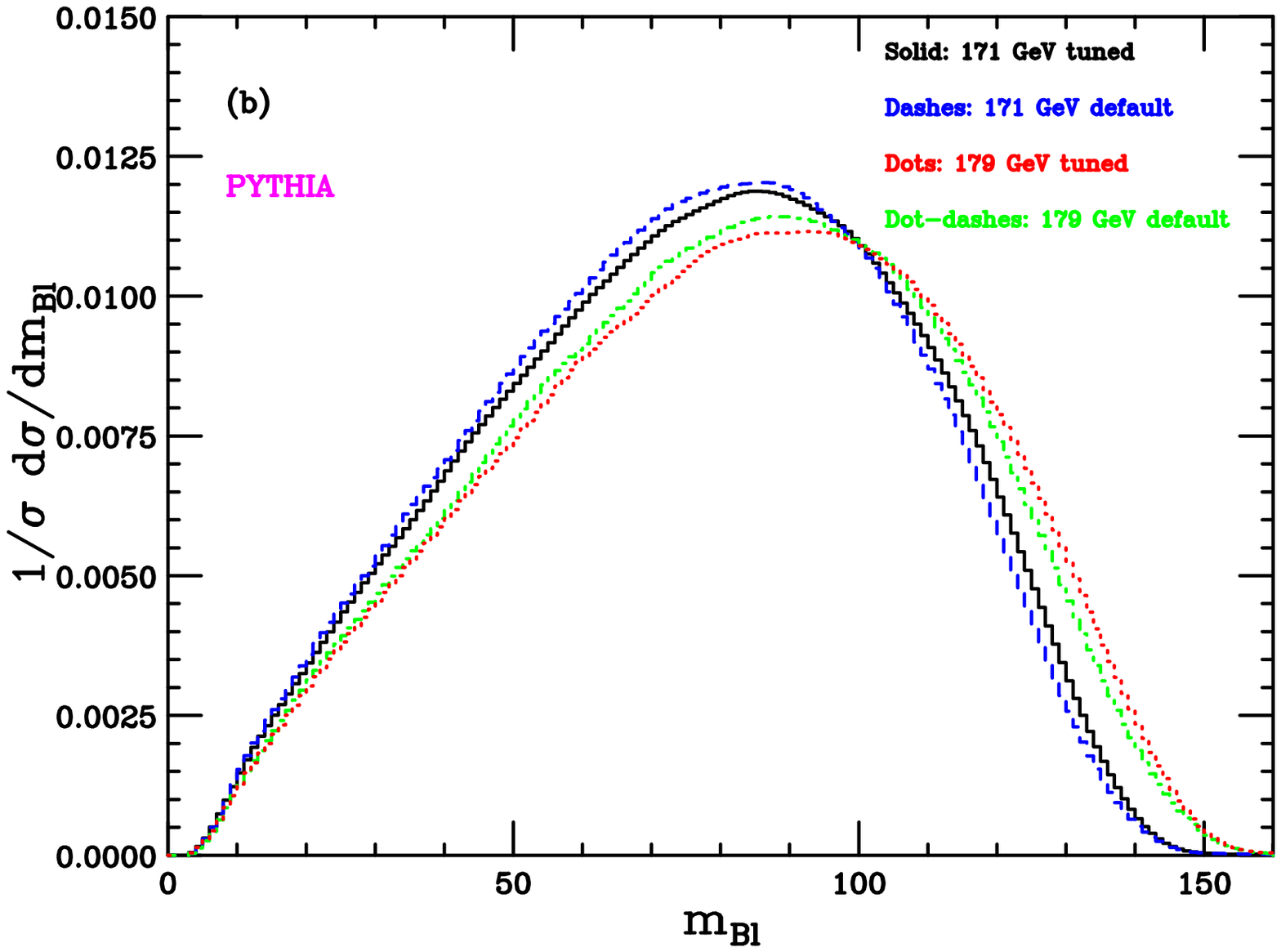}}}
\caption{$B$-lepton invariant mass distribution in top decay, according
to HERWIG (a) and PYTHIA (b), using default and tuned versions,
for $m_t=171$ and 179 GeV.} 
\label{mbl}
\end{figure}
\begin{table}[b]\small
\begin{center}
\begin{tabular}{|c||c|c|c|c|}\hline
$m_t$ (GeV) & $\langle m_{B\ell}\rangle$ (GeV)& 
$\langle m_{B\ell}^2\rangle$ (GeV$^2$)&
$\langle m_{B\ell}^3\rangle$ (GeV$^3$)& $\langle m_{B\ell}^4\rangle$ 
(GeV$^4$)\\
\hline
171 & 78.39 & $7.01\times 10^3$ & $6.82\times 10^5$ & $7.02\times 10^8$ \\
\hline
173 & 79.52 & $7.22\times 10^3$ & $7.12\times 10^5$ & $7.43\times 10^8$ \\ 
\hline
175 & 80.82 & $7.45\times 10^3$ & $7.46\times 10^5$ & $7.91\times 10^8$ \\
\hline
177 & 82.02 & $7.67\times 10^3$ & $7.79\times 10^5$ & $8.37\times 10^8$ \\ 
\hline
179 & 83.21 & $7.89\times 10^3$ & $8.13\times 10^5$ & $8.86\times 10^8$ \\
\hline\end{tabular}
\caption{First four Mellin moments of the $m_{B\ell}$ spectrum according
to HERWIG tuned to LEP and SLD data, as in \cite{drol},
for 171~GeV$<m_t<$~179~GeV.}
\label{mblh}\end{center}
\end{table}
\begin{table}[ht!]\small
\begin{center}
\begin{tabular}{|c||c|c|c|c|}\hline
$m_t$ (GeV) & $\langle m_{B\ell}\rangle$ (GeV)& 
$\langle m_{B\ell}^2\rangle$ (GeV$^2$)&
$\langle m_{B\ell}^3\rangle$ (GeV$^3$)& $\langle m_{B\ell}^4\rangle$ 
(GeV$^4$)\\
\hline
171 & 77.17 & $6.85\times 10^3$ & $6.62\times 10^5$ & $6.81\times 10^8$ \\
\hline
173 & 78.37 & $7.06\times 10^3$ & $6.94\times 10^5$ & $7.23\times 10^8$ \\ 
\hline
175 & 79.55 & $7.27\times 10^3$ & $7.25\times 10^5$ & $7.67\times 10^8$ \\
\hline
177 & 80.70 & $7.48\times 10^3$ & $7.56\times 10^5$ & $8.12\times 10^8$ \\ 
\hline
179 & 81.93 & $7.71\times 10^3$ & $7.91\times 10^5$ & $8.61\times 10^8$ \\
\hline\end{tabular}\caption{As in Table~\ref{mblh}, but employing the
PYTHIA code.}
\label{mblp}\end{center}
\end{table}
\begin{figure}[ht!]
\centerline{\resizebox{0.45\textwidth}{!}{\includegraphics{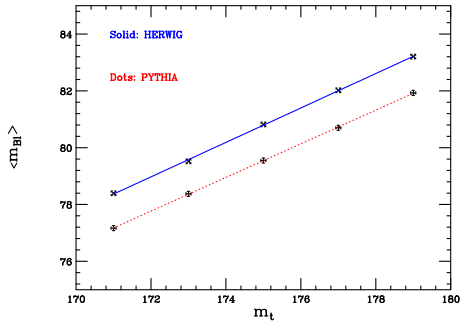}}}
\caption{Average value of the $B\ell$ invariant-mass spectrum in terms
of $m_t$, according to the best-fit straight lines yielded by the tuned versions
of HERWIG and PYTHIA.}
\label{fitt}\end{figure}\par
Fig.~\ref{mbl} presents the $m_{B\ell}$ spectra using tuned and default
versions of HERWIG and PYTHIA, whereas 
Tables~\ref{mblh} and \ref{mblp} quote the first few Mellin moments of $m_{B\ell}$ for
171 GeV~$<m_t<$~179 GeV, according to tuned HERWIG and PYTHIA.
From Fig.~\ref{mbl} one learns that the tuning
has a relevant impact on such spectra; Tables~\ref{mblh} and \ref{mblp}
show that visible differences between HERWIG and PYTHIA are still present,
even after the fits to LEP and SLD data. To relate such a discrepancy  
to an uncertainty on the Monte Carlo top mass, one can try to express 
$\langle m_{B\ell}\rangle$, or even the higher moments, in terms
of $m_t$ by means of a linear fit using the least-square methods.
The best-fit straight lines read:
\begin{eqnarray}
\label{eq1}
\langle m_{B\ell}\rangle_{\mathrm{H}} &\simeq &  
-25.31~\mathrm{GeV} +0.61\  m_t\ \  ;\  \ \delta = 0.043~\mathrm{GeV},\\
\langle m_{B\ell}\rangle_{\mathrm{P}} &\simeq &  
-24.11~\mathrm{GeV} +0.59\  m_t\ \  ;\  \ \delta = 0.022~\mathrm{GeV},
\label{eq2}
\end{eqnarray}
where the subscripts $H$ and $P$ refer to HERWIG and PYTHIA and
$\delta$ is the mean squared deviation in the fit.
Eqs.~(\ref{eq1}) and (\ref{eq2}) correspond to the straight lines in 
Fig.~\ref{fitt}: given the slopes of such lines, the difference 
between HERWIG and PYTHIA $\Delta\langle m_{B\ell}\rangle\simeq 1.2$~GeV implies
an uncertainty $\Delta m_t\simeq 2$~GeV, if one reconstructed $m_t$ from
a possible measurement of $\langle m_{B\ell}\rangle$.
Ref.~\cite{mescia} also pointed out that if one restricted the analysis
to the range 50~GeV~$<m_{B\ell}<$~120~GeV and computed truncated moments,
the induced uncertainty would be reduced to about $\Delta m_t\simeq 1.5$~GeV.
Since such numbers are comparable with the current error on the
top mass world average, the conclusion of \cite{mescia} is therefore that it
is advisable using the tuned versions of Monte Carlo generators, and ultimately
even the HERWIG++ code \cite{hwpp}, as it fares pretty well with respect to
$B$-hadron data.

Before concluding this paper, it is worthwhile discussing the work
\cite{mike}, wherein the top mass is reconstructed as the $W+b$-jet combination
in the lepton+jets channel, using $k_T$ and cone (PxCone, infrared-safe)
algorithms. It was found out that the spectra obtained by using
the $k_T$ algorithm are mostly affected by initial-state radiation and
underlying event, whereas the cone algorithm turns out to be sensitive,
above all, to final-state radiation and hadronization.
The recommendation of \cite{mike} is that using both algorithms will be 
compelling
for the sake of a reliable estimation of the theoretical uncertainty.

In summary, I discussed a few topics relevant to the top mass
reconstruction at hadron colliders, taking particular care about the
relation with the theoretical mass definitions and the uncertainties on
the top mass determination.
In particular, I presented recent work aimed at extracting the pole or 
$\overline{\rm{MS}}$ masses by comparing cross section measurements
and precise QCD calculations and relating the mass implemented in 
Monte Carlo event generators to the jet mass in the SCET framework,
employing $R$-evolution.
Furthermore, progress has been lately achieved in the treatment of
bottom-quark fragmentation in top decays and Monte Carlo tuning,
as well as in the understanding of the sources of theoretical error,
of both perturbative and non-perturbative origin.

%%%%%%%%%%%%%%%%%%%%%%%%

\end{document}